\documentclass[twocolumn]{aastex62}

\graphicspath{{./}{figures/}}

\accepted{\today  ~to ApJ} 

\shorttitle{Stellar Mass Growth Histories in Phase Space}
\shortauthors{Smith et al.}

\begin{document}

\title{Investigating the Stellar Mass Growth Histories of Satellite Galaxies as a Function of Infall Time using Phase-Space}

\correspondingauthor{Rory Smith}
\email{rory.smith@kasi.re.kr}

\author{Rory Smith}
\affil{Korea Astronomy and Space Science Institute (KASI), 776 Daedeokdae-ro, Yuseong-gu, Daejeon 34055, Korea }

\author{Camilla Pacifici}
\affiliation{Space Telescope Science Institute, Baltimore, MD, USA}

\author{Anna Pasquali}
\affiliation{Astronomisches Rechen-Institut, Zentrum f\"ur Astronomie der Universit\"at Heidelberg}

\author{Paula Calder\'on-Castillo}
\affiliation{Departamento de Astronom\'ia, Universidad de Concepci\'on, Casilla 160-C, Concepci\'on, Chile}



\begin{abstract}

	We compile a large sample of nearby galaxies that are satellites of hosts using a well known SDSS group catalogue. From this sample, we create an `ancient infallers'  and `recent infallers' subsample, based on the mean infall time predicted from cosmological simulations for galaxies with their location in phase-space. We compare the stellar mass growth histories of the galaxies in these two subsamples, as determined from multi-wavelength SED fitting that uses a comprehensive library of star formation history shapes derived from cosmological simulations. By simultaneously controlling for satellite stellar mass, and host halo mass, we can clearly see the impact of time spent in their hosts. As we might predict, the ancient infaller population shows clear signs of earlier quenching, especially for lower mass satellites in more massive hosts. More importantly, we find the effects are not limited to massive hosts. We find that hosts which might be considered low mass groups (halo masses $\sim$10$^{13}$~M$_\odot$) can significantly alter their satellites, even for massive satellites (stellar masses $\sim$10$^{10}$~M$_\odot$). Intriguingly, we see changes in the mass growth history of the satellites of clusters as early as 8 or 9 Gyr ago, when they had not yet entered the virial radius of their current host. We propose that this could be the result of galaxies being pre-processed in low-mass substructures in the protocluster outskirts, prior to infall.

\end{abstract}


\section{Introduction} \label{sec:intro}

It is well known that galaxy mass is a key parameter dictating galaxy evolution, but it has now been shown that environmental density also plays an important additional role that is independent from galaxy mass (cf. \citet{Pasquali2010} and \citealp{Peng2010}). However, environmental density cannot tell the whole story as some galaxies may only have recently arrived in their current host, while others may have infallen many gigayears ago and had greater opportunity to be affected by environmental mechanisms such as ram pressure \citep{GunnGott1972}, starvation \citep{Larson1980}, harassment \citep{Moore1996}, and cluster tides \citep{Byrd1990}. 

\begin{figure*}
\includegraphics[width=175mm]{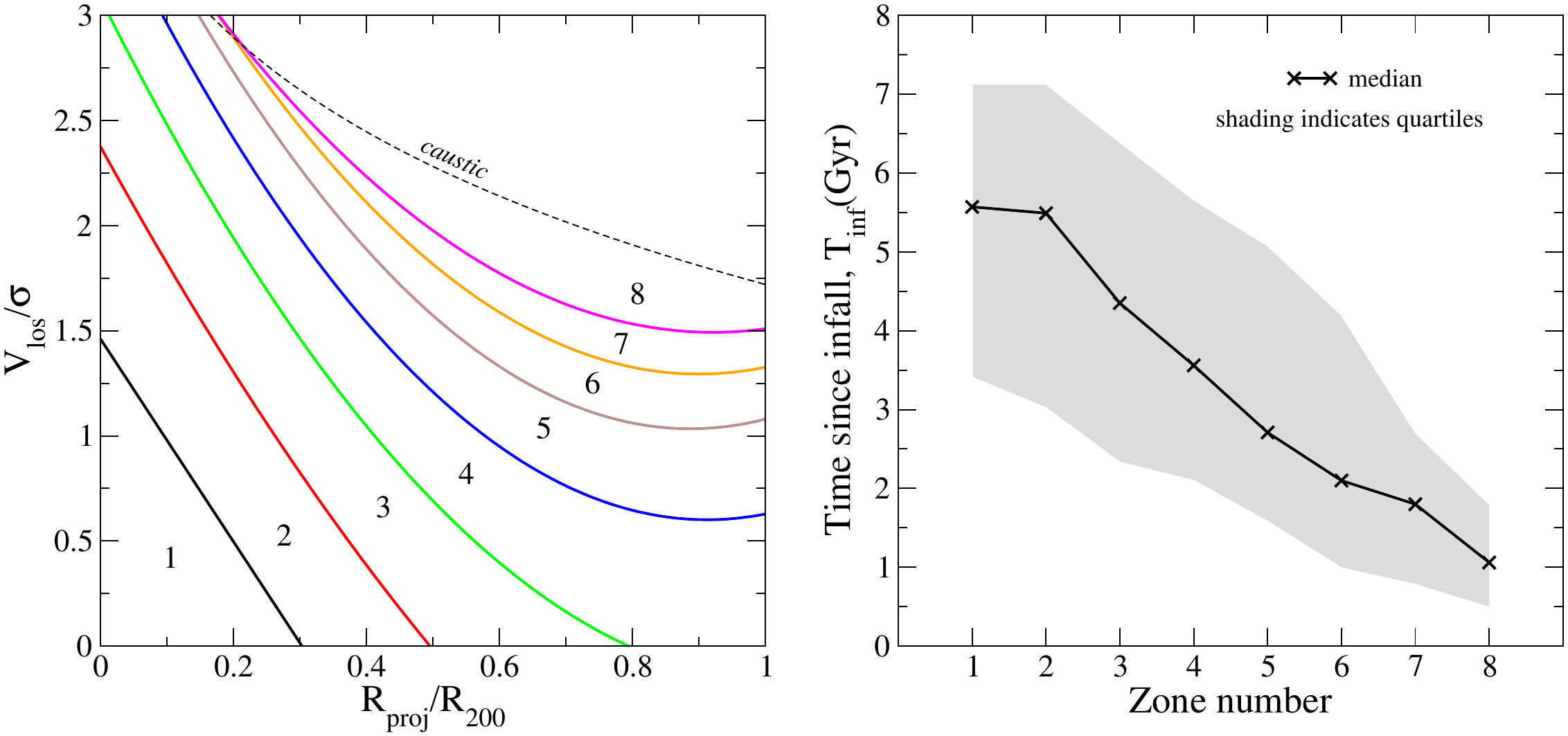}
\caption{{\it Left panel}: The zones defined in \citet{Pasquali2019} of a projected phase-space diagram. The line-of-sight velocity of a galaxy (y-axis) and its projected distance (x-axis) with respect to its host are normalised by the group velocity dispersion ($\sigma$) and the host halo's size R$_{\rm{200}}$. Each zone is labelled with the zone number from 1 up to 8. The dashed line shows the projected caustic of an NFW halo (concentration = 6) computed using Equation 2 of \cite{Jaffe2015}. {\it Right panel}: The median (black line with crosses) and first and third quartile (grey shading) of the infall time distribution within each zone, extracted from cosmological simulations of groups and clusters (see text for details). By splitting up phase-space in this manner, the average infall time of the galaxies in each zone shifts from ancient infallers to recent infallers with increasing zone number.}
\label{compwithsim}
\end{figure*}

Constraining the time that galaxies have spent in their current day environment in a systematic manner, applicable to statistically large samples of galaxies is not trivial. However, a promising approach involves the use of phase-space diagrams -- in this case, plotting the projected clustocentric velocities and radii of a large sample of satellites in groups and clusters. Cosmological simulations have demonstrated that different locations in phase-space are dominated by galaxies with different times since infall (e.g. \citealp{Oman2013}, \citealp{Oman2016}; \citealp{Rhee2017}). In fact, numerous authors have tried to apply the phase-space tool to try to better understand environmental effects (\citealp{Mahajan2011}; \citealp{Muzzin2014}; \citealp{HernandezFernandez2014}; \citealp{Jaffe2015}, \citealp{Oman2016}; \citealp{Yoon2017}; \citealp{Jaffe2018} and many others).

In this study, we investigate how the stellar populations of satellite galaxies change if we compare a sample of galaxies with a mean infall time that was quite recent compared to long ago. To do so, we follow the procedure first described in \citet{Pasquali2019}. The phase-space diagram is split up into 8 zones (see the zones in the left panel of Figure \ref{compwithsim}). The shape of the zones were chosen to approximate contours of mean infall time of dark matter halos from cosmological simulations of groups and clusters seen in projected phase-space (further details below, see also Figure 1 of \citealt{Pasquali2019}). The curve separating each zone has the analytical form: 

\begin{equation}
|\Delta V| / \sigma = a{\rm (R_{\rm proj}/R_{\rm vir})}^2 + b{\rm (R_{\rm proj}/R_{\rm vir})} + c 
\label{zoneeqns}
\end{equation}

\par\noindent
where the coefficients $a$, $b$ and $c$ are expressed as a function of $p$, an integer number running between 1 and 7 as;
\newline

\indent \indent $a = 0.023p^3 - 0.525p^2 + 3.325p - 2.814$, \\ 
\newline
\indent \indent $b = 0.178p^2 - 1.454p - 3.546$, and \\
\newline
\indent \indent $c =-0.105p^2 + 1.229p + 0.338$.
\newline

 We note that the definitions of the coefficients are slightly different from those given in \citet{Pasquali2019}. This is to remove a tiny overlap occurring between the $p$ = 6 and 7 curves. However, the new coefficients produce very similar curves as previously, and would have no impact on our results or the \citet{Pasquali2019} results. We emphasise that the use of these zones should be restricted to within a window from $R_{\rm proj}/R_{\rm vir} = 0$ to 1, and $|\Delta V| / \sigma$ = 0 to 3 (i.e, within the axes range shown in the left panel of Figure \ref{compwithsim}).
 
 In \citet{Pasquali2019}, a suite of 15 hydrodynamical cosmological zoom-in simulations of clusters and groups (the YZICs simulations; \citealp{Choi2017}), is used to study how the time since infall of the galaxy population changes with zone number. Using mock projected phase-space diagrams from the simulations it is shown that when shifting from halos found in zone 1 to halos found in zone 8, the average infall time of the galaxies in each zone systematically shifts from high values to low values with increasing zone number (see right panel of Figure \ref{compwithsim}). In this case, the infall time is defined as the time since a galaxy first crossed the virial radius of the cluster. We emphasise that within each zone there is actually a range of infall times, in part due to projection effects (see brown shading which shows the first and third quartile of the infall time distribution). However, despite the spread in infall times, there is a clear shift in the average infall time with zone number. Thus, by identifying in which zone observed galaxies fall in phase-space, we can test how the observed galaxy properties change as a function of changing mean infall time of the galaxy population.

We successfully applied this approach in \citet{Pasquali2019} to consider how the average luminosity-weighted age, specific star formation rate, metallicity and alpha-abundance of galaxies in each zone change as a function of average time since infall into their host. Here we extend on that approach to consider how infall time might impact on the stellar mass growth history of galaxies in our sample. This is accomplished by cross-matching the SDSS DR7 group catalogue of \citet{Wang2015} with stellar mass growth histories derived from the sample of \citet{Pacifici2016} using SED fitting. Thus one difference is that our galaxy stellar mass growth histories are derived from aperture photometry, where as the galaxy properties in \citet{Pasquali2019} were based on Sloan fiber spectroscopy. In principle, this might mean we are more sensitive to environmental effects that occur from the outside of a galaxy inwards (e.g. ram pressure). 

There is a well known anticorrelation betweeen specific star formation rate (sSFR) and galaxy number density in low redshift galaxies (\citealp{Popesso2011}). Some studies suggest that cluster members have reduced sSFRs compared to field galaxies at redshifts as early as $z$ = 1.4, but the difference decreases with increasing redshift \citep{Alberts2014}. However, the challenges of conducting $z>$ 1 observations means it is difficult to build large galaxy samples, to sample a wide range of environmental densities, or to control for galaxy mass. Therefore, we do not yet have a conclusive picture of environmental impacts at such redshifts, with some conflicting reports in the literature (e.g. \citealp{Elbaz2007}; \citealp{Hayashi2010}; \citealp{Hwang2010}; \citealp{Muzzin2012}).  In contrast to these studies, we use a low redshift sample of galaxies, and attempt to learn about the early cluster environment from their stellar populations today. By combining this with a phase-space analysis, we are also able to consider how their star formation evolves both prior to infall (in the outskirts of their host) and later, after crossing their host's virial radius, and often at times when the host was in an early proto-state, and only a fraction of its current day mass.

In Section 2 we describe the sample of galaxies in groups, and we describe their SED fitting procedure. In Section 3 we present our results. In Section 4 we summarise our results and draw conclusions.

Throughout the Paper, we adopt a Chabrier initial mass function (IMF, \citealp{Chabrier2003}) and a standard $\Lambda$CDM cosmology with $\Omega_{\rm{M}}$=0.3, $\Omega_{\rm{M}}$=0.7 and $h$=0.7. Magnitudes are given in the AB system.

\section{Observational data}
\label{obsdatasect}

\subsection{Sample of galaxies in groups}

\begin{figure}
\includegraphics[width=85mm]{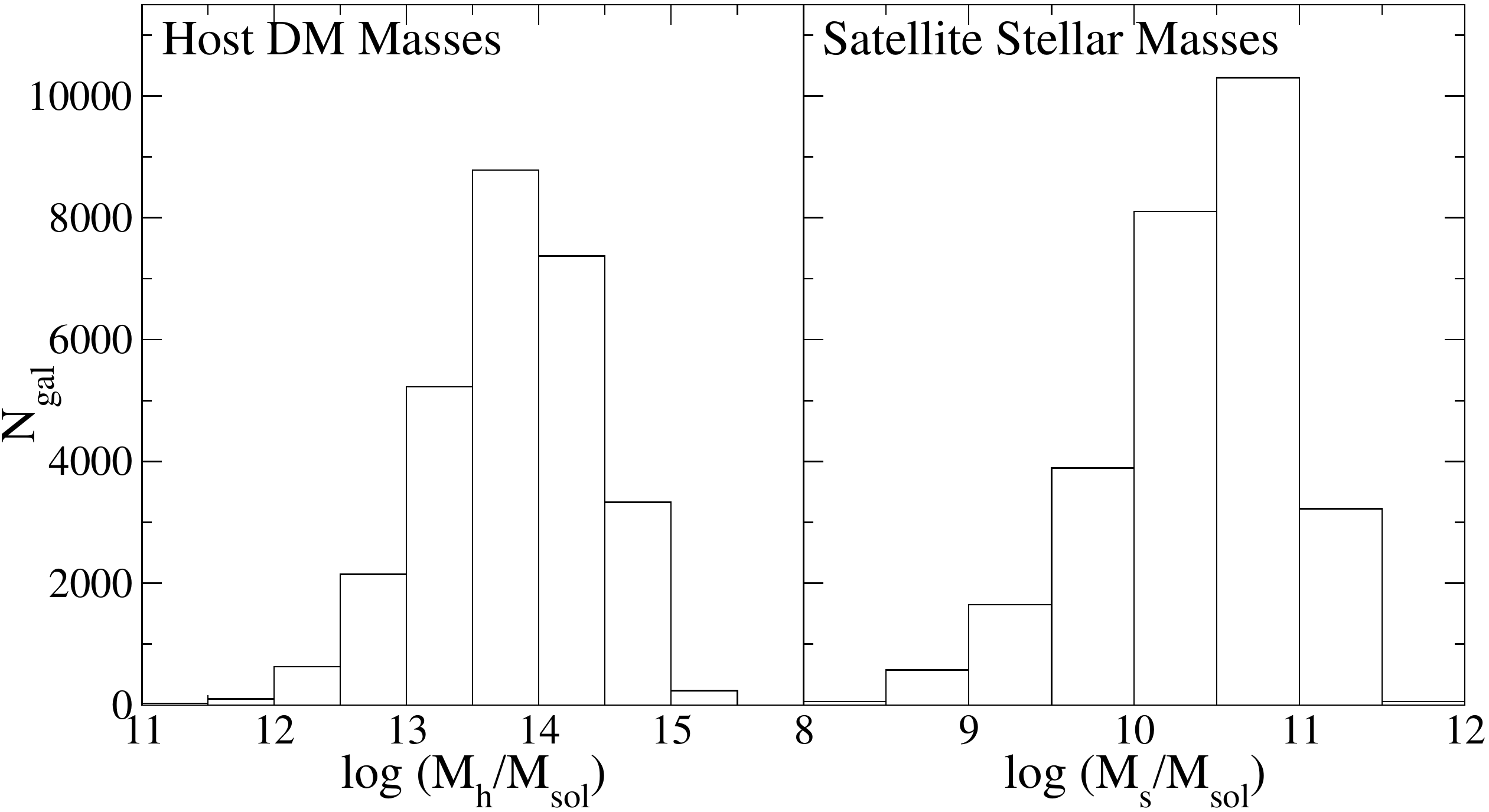}
\caption{Distribution of host masses M$_{\rm{h}}$ (left) and satellite masses M$_{\rm{s}}$ (right) for our complete sample of satellites}
\label{sampledistribs}
\end{figure}

We draw our sample of galaxies from the galaxy group catalogue of \citealp{Wang2015}). These groups were identified in the Main Galaxy Sample of the New York University Value-Added Galaxy Catalogue (\citealp{Blanton2005}) from the SDSS Data Release 7 (\citealp{Abazajian2009}), which have an extinction corrected apparent magnitude brighter than $r$=17.72~mag, and are in the redshift range 0.01 $\le$ $z$ $\le$ 0.20 with a redshift completeness C$_{\rm{z}}>$0.7. For our analysis most of our sample's redshifts are from the SDSS measurements, but with some additional measurements from the literature (sample II of \citealp{Yang2007}). Groups are identified using the adaptive halo-based group finder developed by \citet{Yang2005}. The algorithm uses the traditional Friends-Of-Friends method with small linking lengths to identify candidate groups. Then an iterative approach is followed until convergence to identify group members, and constrain the group mass and size. In the first iteration, the adaptive halo-based group finder applies a constant mass-to-light ratio of 500 $h$  M$_\odot /$L$_\odot$ as a first tentative estimate of the halo mass for each group. This mass is then used to evaluate the size and velocity dispersion of the halo embedding the group, which in turn are utilized to define group membership in redshift space. At this point a new iteration begins, whereby the total enclosed stellar mass is converted into a halo mass based on the halo occupation model of \citet{Yang2005}. This procedure is repeated until the number of group members is fully converged. Further details are provided in \citealp{Yang2007}. The typical uncertainty on M$_{\rm{h}}$ is found to vary between $\sim$0.35 dex in the range M$_{\rm{h}}$ = 10$^{13.5}$ - 10$^{14}$ M$_\odot$/h and $\sim$0.2 dex for halo masses outside this range.

Galaxies are split into group centrals and their satellites. In this study, we only consider the satellite population. We include all groups with 3 or more satellites. However, most have many more satellites. For example, 72$\%$ of our groups have 5 or more satellites, and more than half have 10 or more satellites. This means that most of our hosts really are groups and not simply pairs of galaxies. After applying these various cuts, our sample contains 27,847 satellites, which fall in the redshift range $z$ = 0.018 - 0.160, with roughly two thirds of the sample at $z<$ 0.080. In Figure \ref{sampledistribs}, we show the distribution of host masses the satellites inhabit (left panel) and the distribution of the satellite stellar masses (right panel). These stellar masses are derived from SED fitting as described in Section \ref{SEDfittingsect}. As can be seen in the histograms, the majority of our satellites are in the stellar mass range between 10$^9$-10$^{11.5}$ M$_\odot$ (from massive dwarfs to giant galaxies), and most inhabit hosts with halo masses between 10$^{12}$-10$^{15}$ M$_\odot$ (from giant galaxies to massive clusters). Since our sample is not volume limited, we correct it for Malmquist bias by weighting each satellite by 1/V$_{\rm{max}}$, where V$_{\rm{max}}$ is the comoving volume corresponding to the comoving distance at which that galaxy would still have satisfied the selection criteria of the group catalogue. We will later compare our sample of satellites in groups with field samples (see Section \ref{fieldcompsect})

To plot each of the satellites in a phase-space diagram, like in the left panel of Figure \ref{compwithsim}, we must first move to the frame-of-reference of each host. The group-centric velocity |V$_{\rm{los}}$| of a satellite is its absolute line-of-sight (LOS) velocity subtracted from its host's LOS velocity, where the latter is defined as the mean velocity of all the satellites of that host. The satellite's projected radius (R$_{\rm{proj}}$) is measured with respect to the luminosity-weighted centre of the host. In order to allow us to plot the satellites of many different hosts with many different masses and sizes on a single phase-space diagram, the y-axis is then normalised by the velocity dispersion of each host, $\sigma$, which is the standard deviation of LOS velocities for all the group members. Meanwhile, the x-axis is normalised by R$_{\rm{200}}$ of the host, the radius containing 200 times the critical density of the Universe at that redshift (see Equation 1 of \citealp{Pasquali2019}). We restrict ourselves to only consider satellites that fall within 1 R$_{\rm{200}}$ of the host. By excluding galaxies that are currently outside the cluster, we increase the likelihood of detecting environmental effects. We also ensure that the cluster population dominates over interlopers -- galaxies that are far from the cluster in distance and/or velocity but only appear close by chance due to projection effects. In this sample, 89$\%$ of the sample is in zone $\le$ 4 in which the interloper fraction is $<$ 22$\%$. 

\begin{figure}
\includegraphics[width=85mm]{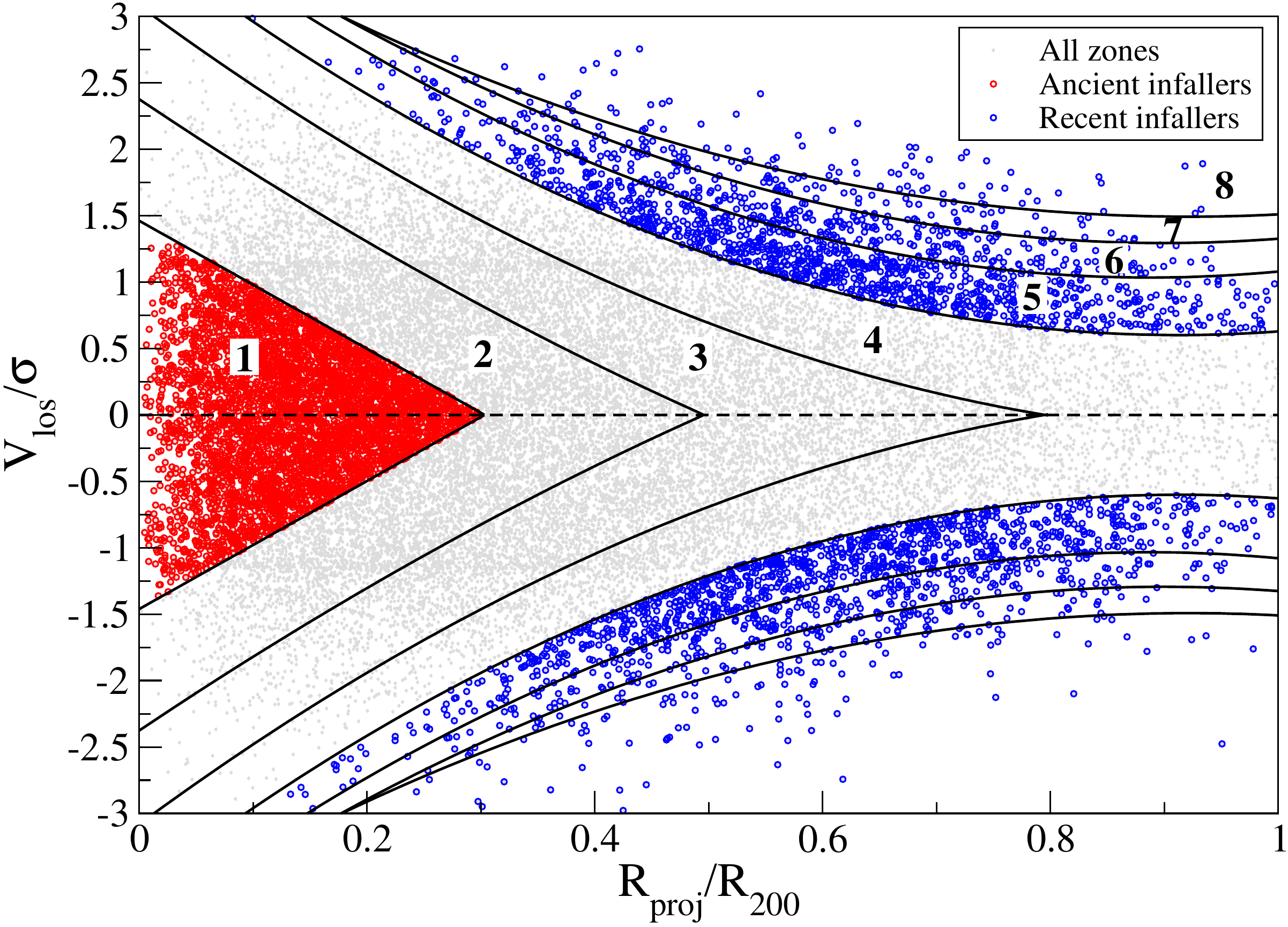}
\caption{The sample in projected phase-space. Grey points are the total sample, red circles are the ancient infaller subsample (zone $=$ 1), and blue circles are the recent infaller subsample (zone $\ge$ 5). The zones (whose boundaries are given by Equation \ref{zoneeqns}) are shown by smooth curves, and labelled with their zone number.}
\label{sampinPS}
\end{figure}

In Figure \ref{sampinPS}, we plot all of our satellites (grey points) in a single projected phase-space diagram. We separate them into the different zones from \citealp{Pasquali2019} (numbered from 1 to 8). We then create a subsample of the galaxies with zones $=$ 1 which we herein refer to as the `ancient infallers' (red points in Figure \ref{sampinPS}). The name comes from \citet{Pasquali2019} where they show using cosmological simulations that galaxies in this region of projected phase-space have a mean time since infall of $\sim5.5$~Gyr (see Figure \ref{compwithsim}). For comparison with this sample, we also create a sample of the galaxies with zones $\ge$5 which we herein refer to as `recent infallers' (blue points in Figure \ref{sampinPS}) as the simulations suggest that this galaxy population has a mean time since infall of $\sim$1 - 2.5~Gyr. 

We caution that, in part due to projection effects, there is actually a range of infall times in any one zone (as illustrated by the shaded quartiles in the right panel of Figure \ref{compwithsim}). Therefore not all of the `ancient infallers' necessarily had an ancient infall. But the main point is that the mean of the distribution of infall times is typically $\sim$4~Gyr earlier for the `ancient infallers' than that of the `recent infallers' (see slope of line in the right panel of Figure \ref{compwithsim}). Therefore, by comparing these two samples, we can study how the mean stellar mass growth history of the galaxy population changes when their mean infall time changes. As such, this approach is best applied to a statistically large sample of satellites, where the population of galaxies in each region of phase-space well samples the distribution of infall times. It is also highly preferable to apply this approach to a sample from multiple groups or clusters, stacked into a single projected phase-space diagram. This is because, in the process of stacking, some of the non-sphericity found in the spatial distribution of galaxies in individual clusters  (e.g. substructure or triaxial shapes) is averaged out.

\subsection{SED fitting method}
\label{SEDfittingsect}
In \citet{Pacifici2016}, the stellar masses and star formation histories (SFHs) of a very large sample of galaxies ($\sim$230,000) at $z<$ 0.16 were constrained using SED fitting. The Pacifici catalogue also provides constraints on the stellar mass growth history of each galaxy which is central to this study, and so we cross-match our sample of satellites with this catalogue. Below we provide a brief description of the SED fitting method but we urge the readers to consider \citet{Pacifici2016} for a more complete description and associated testing.

\begin{figure}
\includegraphics[width=85mm]{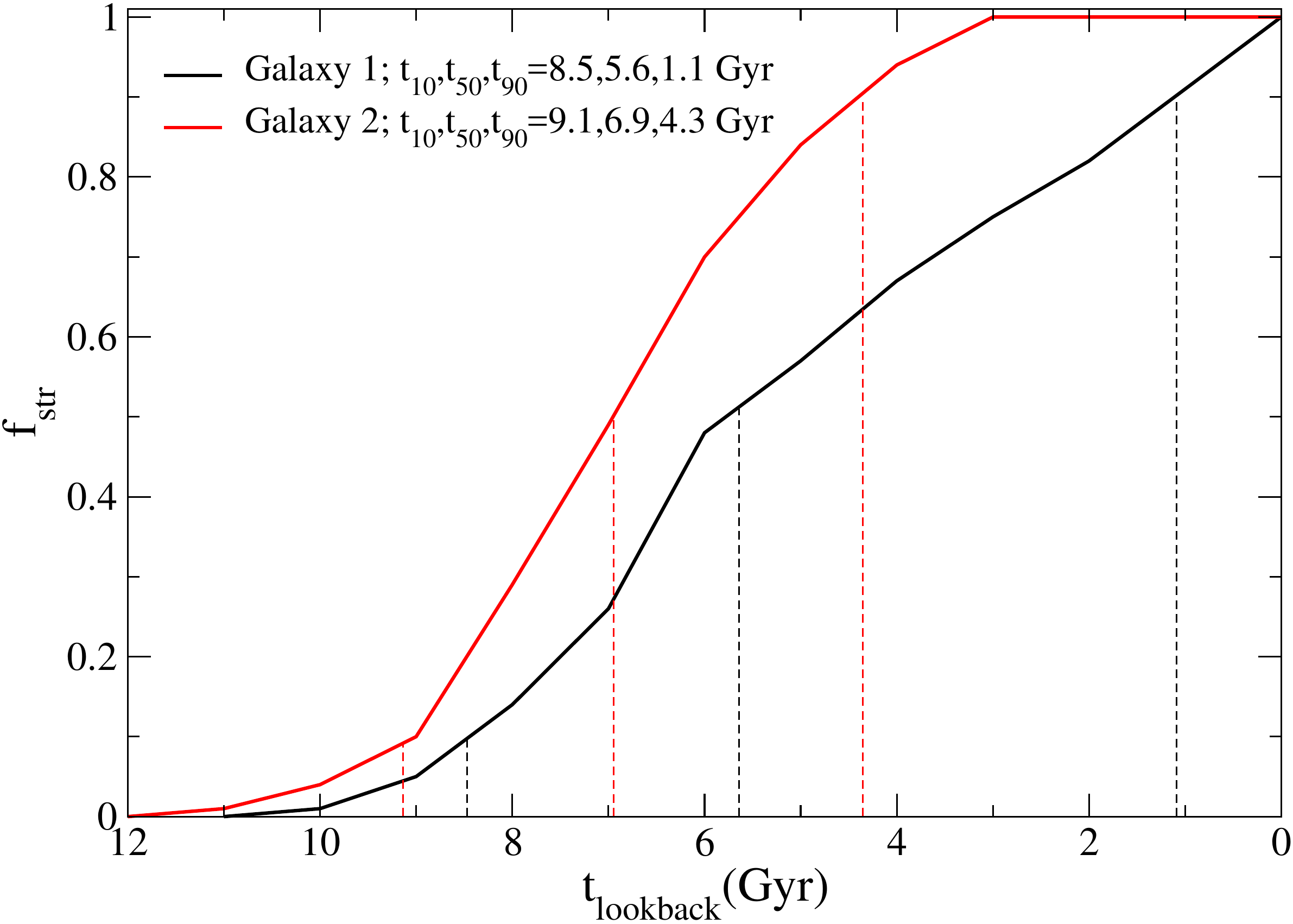}
\caption{Normalised stellar mass growth evolution as a function of look-back time for two galaxies, Galaxy 1 (black lines) and Galaxy 2 (red lines), from our group sample. $t_{10}$, $t_{50}$, and $t_{90}$ (the lookback time when a galaxy forms 10, 50 and 90$\%$ of the total stellar mass formed in its lifetime) are indicated by vertical dashed lines, and the values are given in the legend. The two galaxies are randomly chosen. Galaxy 1 is a recent infaller, low-mass satellite that falls into a low-mass host only recently. Galaxy 2 is an ancient infaller, high-mass satellite that falls into a cluster-mass host.}
\label{mstrevol}
\end{figure}

In \citet{Pacifici2016}, SED fitting is conducted on multi-wavelength broadband photometry spanning from the ultraviolet to near infrared wavelengths. For optical photometry, four broad band filters ($griz$) of the SDSS DR10 (\citealp{Ahn2014}) are used, with a luminosity cut at $r$ $<$ 22.2. Far-ultraviolet (FUV)  and near-ultraviolet (NUV) photometry is from GALEX (Galaxy Evolution Explorer, General Release  6/7; \citet{Martin2005}. 3.4 $\mu$m mid-infrared photometry (W1 band) is provided by the Wide-field Infrared Survey Explorer (WISE; \citealp{Wright2010}), using the aperture correction applied in \citet{Chang2015}. A matching radius of 3 arcsec is applied between surveys. All magnitudes are corrected for foreground extinction using the dust map of \citet{Schlafly2011}, except W1 where the correction is negligible. Optical fiber spectroscopy from SDSS DR10 provides the galaxy redshift required for SED fitting. The spectroscopy is also used to identify and exclude active galactic nuclei (AGNs) using the emission-line catalogue of \citet{Oh2011} to try to reduce AGN contamination which would otherwise interfere with SED fitting.

\begin{figure}
\includegraphics[width=87mm]{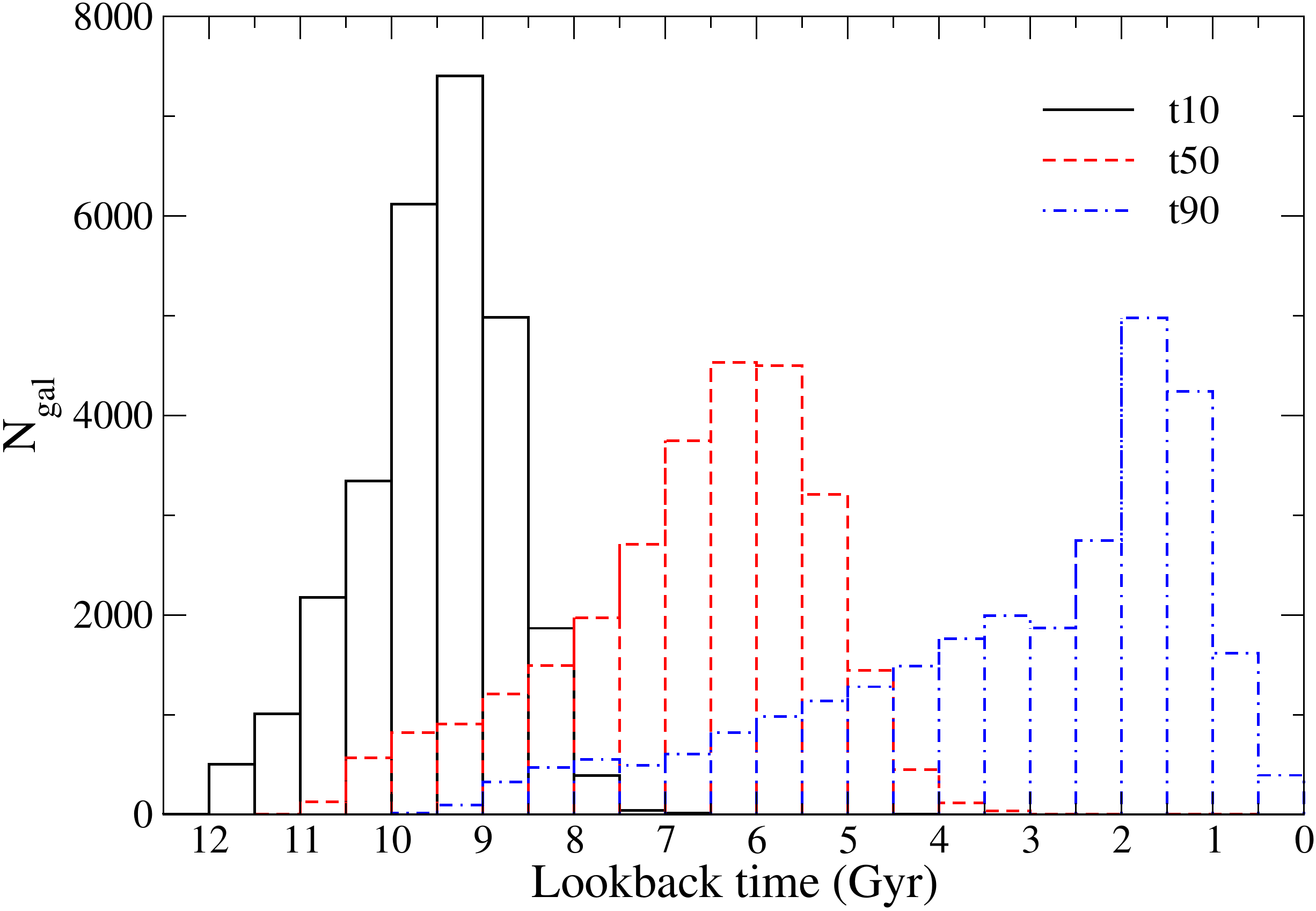}
\caption{Distributions of $t_{10}$, $t_{50}$, and $t_{90}$ (the lookback time when a galaxy forms 10, 50 and 90$\%$ of the total stellar mass formed in its lifetime) for the total group sample}
\label{tvaldistrib}
\end{figure}

Deriving the SFHs of galaxies from multi-wavelength photometry requires a comprehensive spectral modelling technique, and reliable constraints can only be obtained when the library model of SEDs can properly reproduce the data and includes a large variety of SFH shapes. Therefore, the approach used to build the library is to include realistic SFHs from cosmological simulations, stellar and nebular emission are computed consistently, and a detailed model of dust attentuation is included. Realistic and varied SFHs (including declining, rising, roughly constant, bursty and smooth evolution) and metal enrichment histories are extracted from a post-treatment of the Millenium simulation (\citealp{Springel2005}) using the semi-analytical models of \citet{DeLucia2007}. 1.5 million SFHs are generated. These SFHs are then combined with the \citet{Bruzual2003} stellar population synthesis models, and then nebular emission is computed using the photoionisation code {\sc{Cloudy}} \citep{Ferland1996}. Dust attenuation is computed using a more comprehensive version of the \citet{Charlot2000} two component dust model, accounting for uncertainties in the spatial distribution of the dust and orientation of the galaxy \citep{Pacifici2012}.

Now, a Bayesian fitting approach is applied to compare observed SEDs to the model spectral library. In practice, a likelihood that one of the 1.5 million model galaxies reproduces the observed SED is computed. In this way, PDFs of stellar mass, SFR, optical dust, and dust corrected colours are built up for each galaxy. For reasons of computational efficiency, the best estimate SFH is extracted by averaging just the first 10 best fit model SFHs weighted by their likelihood, but tests in \citep{Pacifici2012} demonstrate that the results are negligibly changed if the entire library is used instead. We note that the fitting is made only with the shape of the SFH, and not with the normalisation. Instead when fitting, all model galaxies are normalised to the luminosity of the observed galaxy.

Finally, the SFHs are parameterised using $t_{10}$, $t_{50}$, and $t_{90}$, the lookback time when a galaxy forms 10, 50 and 90$\%$, respectively, of the total stellar mass formed in its lifetime. There are mean uncertainties of 0.93~Gyr, 0.77~Gyr, and 0.32~Gyr on $t_{10}$, $t_{50}$, and $t_{90}$, respectively. These parameters are crucial to this study. The fractional mass growth history of two real examples of galaxies from our sample are shown in Figure \ref{mstrevol}. Galaxy 1 shows a quite constant gradient, which demonstrates it has steadily increased its stellar mass at a fairly constant rate over its history, and the assembly likely continues until today. Meanwhile, Galaxy 2 has assembled its stellar mass more rapidly that Galaxy 1 at higher redshift, and reached its final mass at a lookback time of approximately 3~Gyr, and has since not assembled any new stellar mass. The full distribution of $t_{10}$, $t_{50}$, and $t_{90}$ for our sample of satellites is shown in Figure \ref{tvaldistrib}.

\section{Results}

\subsection{Approach}
As described in Section \ref{obsdatasect}, we have compiled several key parameters for each of the satellites in our sample. From the location in phase-space, we can derive the zone number (labelled 1-8 in Figure \ref{compwithsim}) which is a proxy for the mean infall time of galaxies in that zone. From the group catalogue, we have the halo mass of a satellite's host (M$_{\rm{h}}$). From SED fitting, we know the satellite's stellar mass (M$_{\rm{s}}$), and have a constraint on the stellar mass growth history from the parameters $t_{10}$, $t_{50}$, and $t_{90}$, the lookback time when a galaxy forms 10, 50 and 90$\%$ of the total stellar mass formed in its lifetime. 

In this study, our primary interest is to understand how the mean time since infall of a population of satellites into their host can alter their stellar mass growth history. However, to more clearly see the effect of the time since infall, it is necessary to first control for both satellite mass and host mass. This is because these two parameters will both independently alter the stellar mass growth history. For example, at fixed satellite mass, more massive hosts like clusters might be expected to be more efficient at quenching galaxies than lower mass hosts. Similarly, it is well known that more massive galaxies quench their star formation earlier, even in the absence of a host environment -- a process referred to as `mass quenching' \citep{Birnboim2007}.

Therefore, our approach is to simultaneously control for both the host mass and satellite mass, and then compare the `ancient infaller' sample (zone $=$ 1) to the `recent infaller' sample (zone $\ge$ 5) and look for the impact on the stellar mass growth history in terms of the parameters $t_{10}$, $t_{50}$, and $t_{90}$. 

\begin{figure*}
\includegraphics[width=170mm]{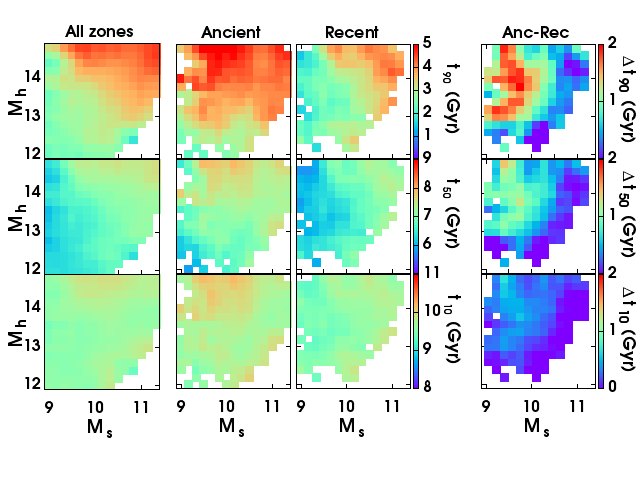}
\caption{In these panels, we present how changing the mean infall time of the satellite population into their host has impacted on their satellites mass growth history (parameterised through $t_{90}$, $t_{50}$, and $t_{10}$). The left column shows the full sample of satellites (labelled `All zones'). Satellites in the second column can be considered ancient infallers (zones $=$ 1; labelled `Ancient'). Satellites in the third column can be considered recent infallers (zones $\ge$ 5; labelled `Recent'). The fourth column shows the pixel-by-pixel subtraction of the `Ancient' and `Recent' panel (labelled `Anc-Rec'). Each panel has the current host mass (M$_{\rm{h}}$) on the y-axis versus the satellite's current stellar mass (M$_{\rm{s}}$) on the x-axis. Each pixel is colour-coded by mean  $t_{90}$ (top row), $t_{50}$ (middle row), and $t_{10}$ (bottom row). Please note the changing range of the colour bars between rows, except in the last column. We note that the colour bar range remains larger than the uncertainties in their corresponding $t$ parameter. To reduce noise from low statistics, pixels are shown blank if they contain less than 4 satellites.}
\label{tvalfig}
\end{figure*}

\subsection{Effect of infall time on stellar mass growth of satellites as measured from $t_{10}$, $t_{50}$, and $t_{90}$}

In Figure \ref{tvalfig}, each panel has host mass (M$_{\rm{h}}$) on the y-axis versus satellite's stellar mass (M$_{\rm{s}}$) on the x-axis. The colour of a pixel shows the mean value of $t_{90}$ (top row), $t_{50}$ (middle row), and $t_{10}$ (bottom row) of galaxies falling in or near that pixel. In practice, we calculate the mean including all galaxies that fall within a 1 pixel wide border surrounding the central pixel. This partly smooths the data, in order to bring out the large scale trends across the panel. The left column is labelled `All zones' meaning it contains the complete sample of satellites. The central column is the `ancient infallers' subsample (zones $=$ 1), and the right column is the `recent infallers' subsample (zones $\ge$ 5).

We first concentrate on the $t_{90}$ panels (top row), starting with the total sample (column labelled `All zones'). By definition, $t_{90}$ is when the majority (90$\%$) of the stellar mass of the galaxy was assembled. Thus galaxies that stop forming stars a long time ago must have large $t_{90}$ (red pixels), while galaxies that continue to form stars today have small $t_{90}$ (blue pixels). A blue-to-red colour gradient can be seen with increasing satellite stellar mass (along the x-axis) and increasing host halo mass (along the y-axis). Thus both increased satellite mass and increased host mass act in parallel to change the mass growth history of satellites (i.e. combined mass and environmental quenching, respectively). This further confirms our need to control for both these parameters simultaneously to more clearly detect the effect of changing the mean infall time of the satellite population. 

We now compare $t_{90}$ for the ancient infallers subsample (column labelled `Ancient'; zones $=$ 1) with the recent infallers subsample (column labelled `Recent'; zones $\ge$ 5). To better highlight differences between the ancient and recent infaller panels, on the far right we show the pixel-by-pixel subtraction of the `Ancient' and `Recent' panels (column labelled `Anc-Rec'). The colour bar of the `Anc-Rec' column differs from the other columns to better emphasise the differences.

By eye, it is clear that there are significant differences between the ancient and recent infallers. The ancient infaller panel has a lot more red pixels meaning that $t_{90}$ occurred earlier for them (i.e. they were quenched longer ago). Compared to the recent infaller panel, the colour gradient is more vertical, meaning the ancient infallers are more sensitive to their host environment. Meanwhile, the recent infaller panel has colour gradient that is quite diagonal, moving from red to blue towards the upper-tight corner of the panel. This shows that the recent infallers care about both their stellar mass and host mass.

The `Anc-Rec' panel reveals that at high stellar masses, the ancient and recent infallers have very similar $t_{90}$. In short, the massive satellites of both the ancient and recent infallers are dominated by mass quenching, with no clear indication of environmental quenching in their $t_{90}$ parameter. However, with decreasing stellar mass combined with increasing host mass, quenching occurs more quickly, with the $t_{90}$ values of the ancient infallers being shifted $\sim$2~Gyr earlier than in the recent infallers. As an indicator of environmental effects, this is physically intuitive. {\it{Lower mass satellites are more susceptible to the environment of more massive hosts, especially for those that spent more time in that environment.}} In this picture, the fact that the recent infallers have a weaker dependence on host mass than the ancient infallers also makes sense -- {\it{many of the recent infallers may simply not have had time to respond since entering their new environment}}.

The `Anc-Rec' panel also reveals another important take away point. Green/orange/red pixels are visible for M$_{\rm{h}}=10^{13}~$M$_\odot$ or lower, and also for M$_{\rm{s}}$ as large as $10^{10}~$M$_\odot$. This means that {\it{even low mass hosts are significantly changing the stellar mass growth of their satellites, and even when their satellites are quite massive galaxies}}.

In both the ancient and recent infaller panel, the distribution of $t_{90}$ values is notably less smooth compared to the `All zones' panel. The ancient infaller and recent infaller subsamples contain only 17$\%$ (4815 satellites) and 11$\%$ (2931 satellites) of the total sample, so the reduced smoothness of the results is partially the result of decreased number statistics. We also try dividing the sample into a zone $<$ 3 subsample and a zone $\ge$ 3 subsample which results in better statistics within the two subsamples (see Appendix, Figure \ref{tvalfig_diffzones}). Qualitatively the results are similar but because the zone $<$ 3 and zone $\ge$ 3 subsamples do not occupy such extreme ends of the range of zone numbers, we do not detect such strong differences between them. As such, we find it is better to compare the ancient and recent infaller subsamples, and we do so for the rest of the paper. Although they may have smaller sample sizes, it is clearly demonstrated in Figure \ref{tvalfig} that there are sufficient statistics to see clear trends with host mass and satellite mass. In particular, the ancient infaller panel is notably more red than the recent infaller panel in the upper-left corner of the two panels.

The $t_{90}$ results (top panels) show the same overall trends as Figure 8 of \citet{Pasquali2019}, which is a plot with the same axes but with data points coloured by luminosity-weighted age instead. However, one important difference between the studies is that \citet{Pasquali2019}  derived galaxy properties from fiber spectroscopy, whereas here we use aperture photometry. Therefore, in principle the aperture photometry might be slightly more sensitive to environmental effects as it is not confined to measuring light from the galaxy centres. Unfortunately, it is difficult to compare the results directly in a quantitative manner as $t_{90}$ and the luminosity-weighted age are fundamentally different quantities. Nevertheless, qualitatively the differences between ancient and recent infallers are broadly similar. 

In our projected phase space diagram, the fraction of interlopers tends to increase with zone number (see Table 2 of \citealt{Pasquali2019}). As a result, the recent infaller sample contains more interlopers which tends to wash out some of their host mass dependency, which in turn leads to an increase in the host mass dependency in the `Anc-Rec' panel. Fortunately, the presence of a clear host mass dependency in the recent infaller $t_{90}$ panel demonstrates the limited impact the interlopers are having on our results.

In the middle row, we consider the $t_{50}$ values of the satellite sample. We first focus on the `All zones' column. It is clear that the general trends seen in the $t_{90}$ panels are still visible. We deliberately vary the colour bar range and midpoint between rows to try to emphasise any trends that exist. The `Anc-Rec' panel shows the differences are not so large in $t_{50}$ compared to $t_{90}$, but the general trend for the largest differences to be found in the upper-left corner of the panels remains.

Nevertheless, for $t_{10}$ in the bottom row, the same trend is not clear in the `All zones' panel. It is only when we split the sample into the ancient and recent infaller subsample, that the differences emerge more clearly. This result highlights how, by taking advantage of knowledge of the location of galaxies in phase-space, we are able to increase our sensitivity to detecting environmentally induced changes. Without knowledge of their infall time, we will always mix galaxies that have been in the environment for a long time and been affected with those that only recently entered the environment, and thus the signal of environmental effects is diluted. The `Anc-Rec' panel shows the $t_{10}$ differences are small ($<$ 1~Gyr) compared to $t_{50}$ and $t_{90}$, but that the same overall trend across the panel exists.

At first glance, it might appear that the differences appearing in the $t_{10}$ panels suggest that the ancient infaller population suffers alterations to their very early star formation history compared to recent infallers. However, we cannot be certain if this is occurring from these plots alone. This is because if a galaxy falls into a cluster and has its star formation quenched at earlier times then, compared to a galaxy that is not quenched, $t_{90}$ will occur earlier and the final stellar mass will be reduced. As $t_{10}$ and $t_{50}$ are measured when the stellar mass reaches a specified fraction of the final stellar mass, the reduced final mass in turn forces both $t_{10}$ and $t_{50}$ to occur earlier. Therefore it is difficult to detect differences in the early star formation history using the $t_{10}$, $t_{50}$, and $t_{90}$ parameters as they are all tied to the final stellar mass of the galaxy which is affected by changes occurring at any time during the mass growth evolution. Furthermore, our results may be impacted by progenitor bias, as we compare galaxies at fixed stellar mass, even though they might have had very different final stellar masses if they hadn't had their star formation changed by the environment. In the following sections, we attempt to reduce the impact of these issues.

\begin{figure*}
\includegraphics[width=180mm]{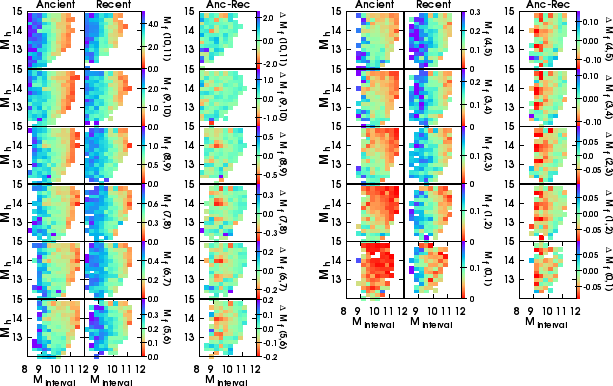}
\caption{At one gigayear intervals of lookback time (where t = 0 corresponds to $z$ = 0), we compare M$_{\rm{f}}$, the fraction of the stellar mass at that interval which is formed during the interval (as indicated on the colour bars) for the ancient infaller vs recent infaller subsamples (see column headers). Each panel has current day host mass M$_{\rm{h}}$ on the y-axis, and the mass of the satellite at the epoch of the interval M$_{\rm{interval}}$ on the x-axis. As a result, a star forming galaxy will move horizontally from left to right between the intervals as the stellar mass grows, but not vertically. The look back time of the interval is indicated in brackets on the colour bar label. The colour bar is linear in all panels, and its range is arbitrarily chosen between intervals to try to highlight differences between the ancient and recent infaller panel. But the range is fixed for both the ancient and recent infaller panel to enable a fair comparison. We also provide the `Anc-Rec' column which is the pixel-by-pixel subtraction of the ancient and recent infaller panels so as the differences can be more clearly seen. The colour bar range is scaled to be $\pm$half the maximum of the colour bar range in the ancient and recent panels. \textit{There are two takeaway messages from this figure; (i) As might be expected, the clearest differences in star formation activity between ancient and recent infallers is seen in the most recent intervals since galaxies fell into their hosts, especially for low mass satellites in more massive hosts. (ii) However, the differences in star formation activity between ancient and recent infallers visibly persist back to intervals as early as [8,9] Gyr ago, several gigayears prior to the mean infall time of the ancient infallers.}}
\label{diffsfig}
\end{figure*}

\subsection{Stellar mass growth in lookback time intervals} 
In Figure \ref{diffsfig}, we consider the stellar mass formation occurring within 1~Gyr time intervals of lookback time (where t=0 corresponds to $z$ = 0), starting from the most ancient [10,11] Gyr ago, and moving through to the most recent [0,1] Gyr ago. In each interval, we compare the stellar mass formed in the ancient infaller (column labelled `Ancient') and in the recent infaller sample (column labelled `Recent'). Comparing these panels allows us to clearly see differences in the mass growth histories of the two samples over the age of the Universe and, to better highlight the differences, we also present the pixel-by-pixel difference between these two panels (column labelled `Anc-Rec'). As in earlier plots, the current host mass M$_{\rm{h}}$ is on the y-axis. However, on the x-axis we use the mass of the galaxy {\it{at that interval of time}}, instead of using the final stellar mass as in previous plots\footnote{In practice, we use the mass at the upper bound of the interval.}. In this way, we can study how galaxies of a particular mass at that epoch were forming stars during that epoch, and look for differences between the ancient and recent infallers. An individual galaxy will move horizontally only, from left to right across the x-axis between intervals, if star formation causes it to grow in mass. 

The colour of a pixel is determined by the amount of stellar mass formed in that interval, normalised by the mass of the galaxy in that interval. In other words, it is a measure of the fraction growth of the galaxy's stellar mass during that interval. For example, a galaxy with M$_{\rm{f}}$=10 formed ten times its initial mass during that one gigayear interval. M$_{\rm{f}}$ could also be considered as a mean specific star formation rate (sSFR), only increased by a factor of 10$^9$ for each year of the gigayear interval. It should be noted that the range of the colour bars decreases substantially as we move from the early Universe to more recent times. This clearly shows that star formation rates were significantly higher at earlier epochs, even at fixed galaxy mass. We choose the colour bar range fairly arbitrarily to try to better highlight the differences between the ancient and recent infaller panel, thus we caution against comparing between intervals (i.e. rows) by eye. However, the ancient and recent infaller panels share the same colour bar, meaning it is possible to compare these two columns by eye in any one interval, and it is this comparison which the figure is designed to focus on. We quite arbitrarily choose the range of the `Anc-Rec' panel colour bar to be $\pm$half the upper limit of the `Ancient' and `Recent' colour bar, to try to better highlight the differences between them.

Starting at the most recent intervals, [0,1] and [1,2] Gyr ago (shown in the bottom-right of Figure \ref{diffsfig}), it is clear that there is a significant difference between the ancient and recent infaller panel. Red pixels (low star formation) cover most of the ancient infaller panel. In comparison, the recent infallers have blue pixels (high star formation), especially for low mass satellites. The red/orange pixels in `Anc-Rec' panel shows that many ancient infallers show reduced stellar mass growth compared to the recent infallers. These red/orange pixels tend to be located in the upper-left of the panel, meaning this preferentially occurs for lower mass satellites in higher mass hosts. But red pixels can be found across a wide range of both host mass and satellite mass. This is similar to what we saw in the $t_{\rm{90}}$ panel of Figure \ref{tvalfig}.

Now, as we move further backwards in time through the intervals, it is clear that the same qualitative differences between the ancient and recent infaller panels can be seen at successively earlier intervals. The recent infaller panel always contains more bluer pixels, and they tend to spread further across to the right of the panel (towards higher satellite masses). The `Anc-Rec' panel continues to show a gradient towards more red pixels in the upper left corner. 

\begin{figure*}
\includegraphics[width=180mm]{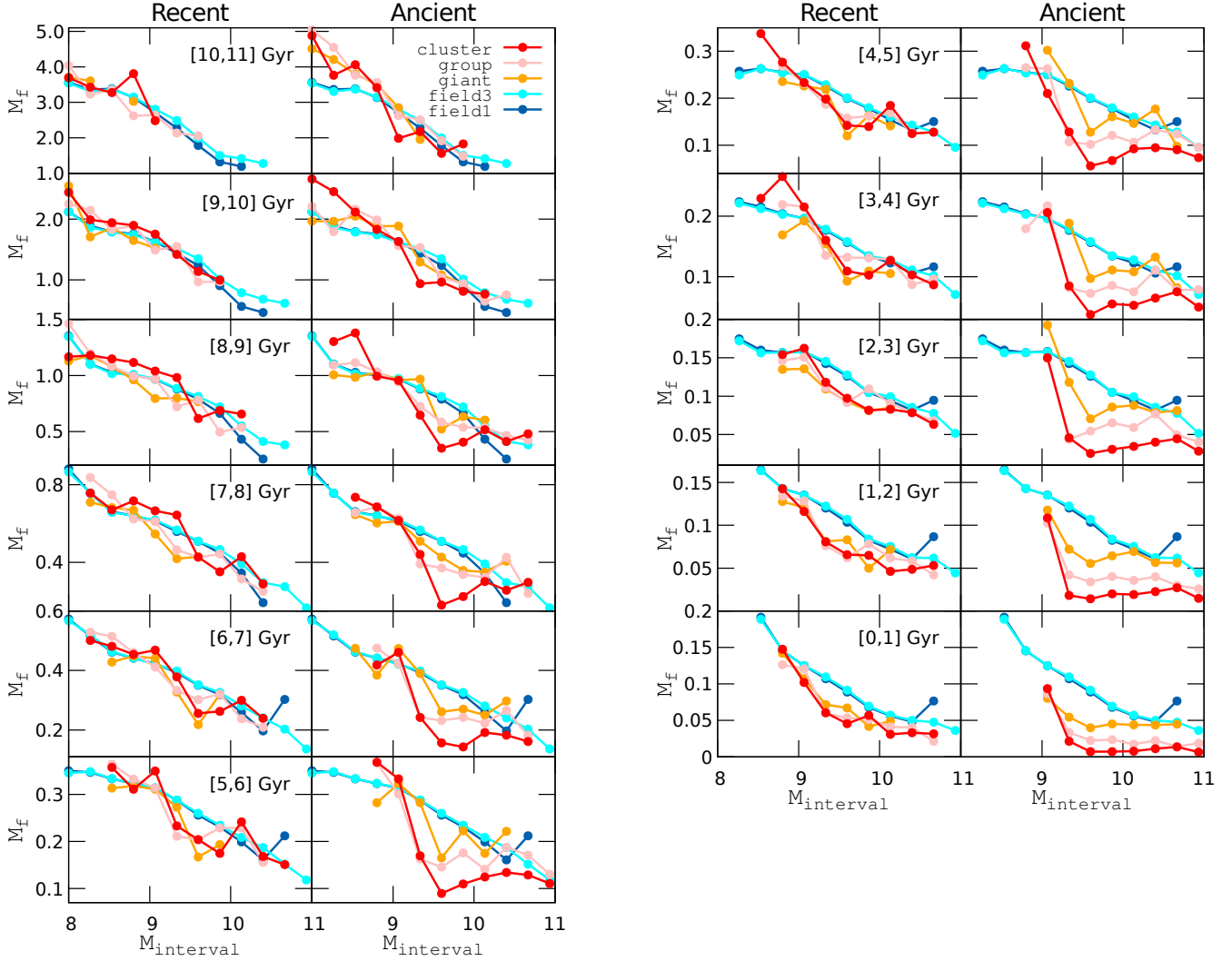}
\caption{At one gigayear intervals of lookback time (interval indicated in brackets in the panel) where t = 0 corresponds to $z$ = 0, we plot M$_{\rm{f}}$ (the fraction of the stellar mass at that interval which is formed during the interval) on the y-axis against M$_{\rm{interval}}$ (the mass of the satellite at the epoch of the interval) on the x-axis for the recent and ancient infallers (see column headers). The environment of the galaxy is shown by the colour of the curve (see legend), and includes cluster, group and giant galaxy mass hosts, and two control field samples (see text for details of definitions). Please note that the y-axis range changes with the lookback time of the interval.}
\label{intSFcurvesfig}
\end{figure*}

 The same trend for more red pixels to appear in the upper left corner is repeated in most of the intervals, although the strength of the difference between the ancient and recent infallers becomes less prominent as we approach the earliest intervals. However, the differences remain quite visible even at intervals as early as [8,9] Gyr ago. This is long before the great majority of galaxies would have entered their hosts. The mean infall times derived from phase-space are measured with respect to the current redshift of the galaxy. Therefore, we also try remaking the plot using intervals of lookback time measured from the current redshift of the galaxy (i.e. t=0 corresponds to the galaxy redshift), and find broadly consistent results (see Appendix, Figure \ref{diffsfig_lookbackint}). The results confirm that differences between the ancient and recent infallers arise at least [8,9] Gyr before entering their hosts. Furthermore, at these times their hosts would be expected to be significantly less massive than they are today and thus probably have a much less dense intracluster medium, and more shallow cluster potential. And yet our results suggest that these galaxies already show indications of suppressed star formation at these early times, despite being beyond their host's virial radius, when compared with recent infallers of equal mass at the same epoch. 

We tentatively propose that these results could be understood in the framework of assembly bias. The fact that recent infallers only recently entered the cluster almost certainly suggests they were initially further from the cluster in these early epochs (e.g. interval [8,9] Gyr ago). It could be that, far from the protocluster, they inhabited lower density environments which somehow promoted early star formation. Perhaps gas supplies were more plentiful, or more easily accreted in this environment. Alternatively, perhaps the ancient infallers began life in the more dense volume nearer to the protocluster, where there was a greater abundance of group mass hosts that would eventually merge together to form the cluster. Given that our results have already highlighted the ability for quite low mass groups to have a significant effect on their satellites mass growth (e.g. Figure \ref{diffsfig}), it is plausible that these early groups could already begin reducing their star formation at these epochs.

\subsection{Comparison with a `field' sample}
\label{fieldcompsect}

We wish to compare our satellite population mass growth history with a sample of galaxies that should be relatively free from environmental effects. We define a `Field 1' and `Field 3' sample, following the nomenclature of \citet{Pasquali2019}. `Field 1' galaxies are defined as any central galaxy residing in hosts less massive than 10$^{12}$~M$_\odot$. `Field 3' galaxies are all central galaxies with no associated satellite (down to the magnitude limit of the SDSS spectroscopy) and that are separated from the nearest host halo by at least 5 virial radii of the nearest host. With these definitions, our field samples should be considered two different ways to define galaxies in much lower density environments than the rest of our group sample, rather than being truly representative of the field environment. There is significant overlap between the two samples, but the field 1 sample central galaxies tend to have $z$=0 stellar masses limited to $\sim$10$^{11}$~M$_\odot$, where as field 3 central extend up to $\sim$10$^{12}$~M$_\odot$.

In Figure \ref{intSFcurvesfig}, we attempt to compare the stellar mass growth in each Gyr interval for our two field samples (blue and cyan curves; see legend) with our satellite sample. We split the satellites by their host masses; cluster, group and giant galaxy (red, pink and orange curves corresponding to $\log$(M$_{\rm{h}}$/M$_\odot$)= 14-15, 13-14, and 12-13, respectively). The x-axis is stellar mass at the time of the interval, and the y-axis is M$_{\rm{f}}$ (the fraction of the galaxies mass at that interval which was produced in new stars during that interval). We show a separate column for the recent and ancient infallers. The field 1 and field 3 curves are simply repeated in both the ancient and recent infaller panels. In this way, we attempt to take the trends visible in the panels of Figure \ref{diffsfig} and convert them into a more graphical form, for ease of comparison with the field samples.

We note that the standard deviation in M$_{\rm{f}}$ about each data point is quite large, typically spanning about half the vertical range of the panel. Nevertheless, despite the large variance between individual galaxies, the clear similarity between the panels of nearby intervals, and the smooth transition of the curves shape with changing interval time, clearly demonstrates the evolving mean properties of the various subsamples.

\begin{figure}
\includegraphics[width=90mm]{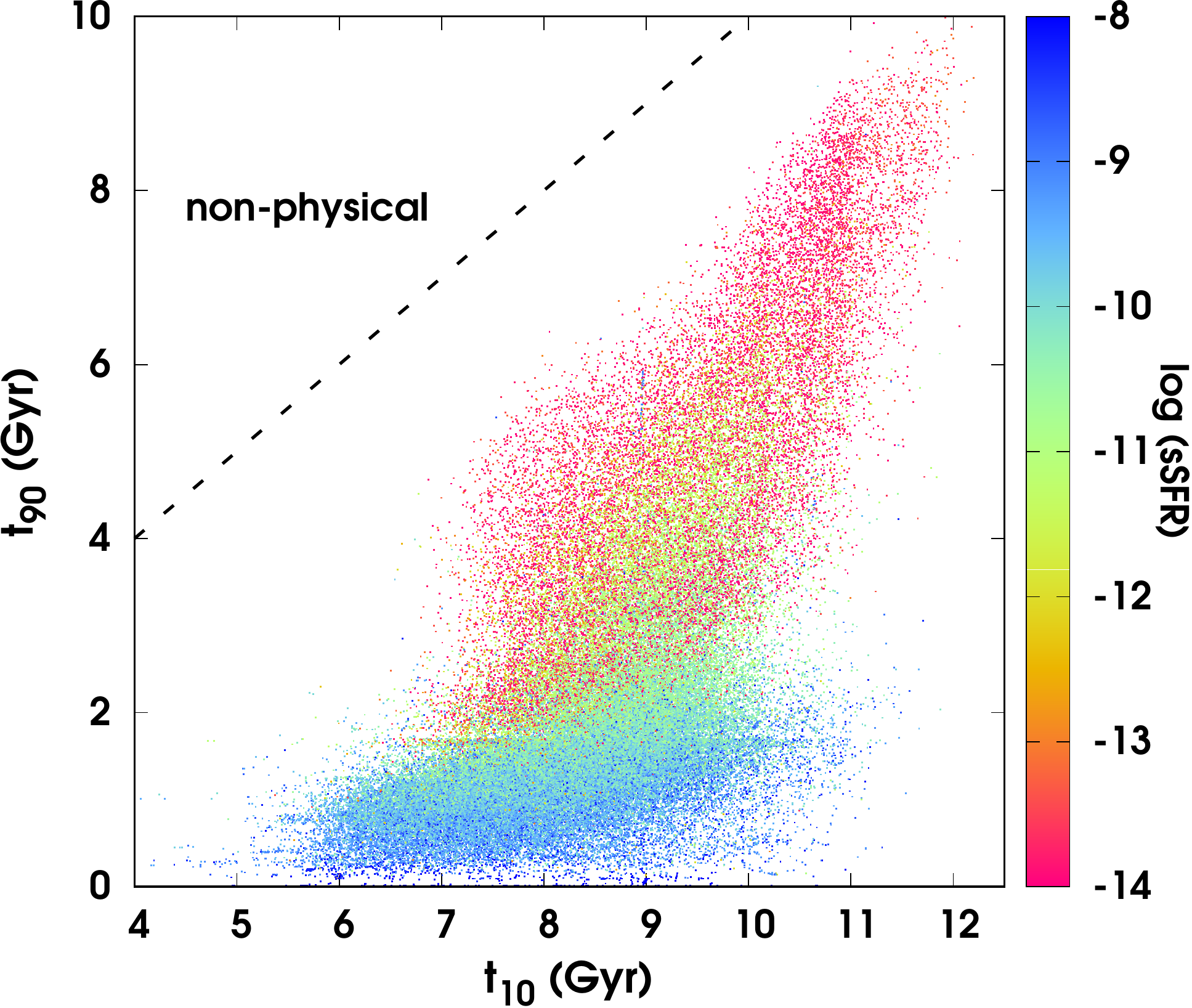}
\caption{$t_{90}$ vs $t_{10}$ for a control sample of field galaxies. Data points are coloured by current specific star formation rate. A clear trend between the two parameters exists, as would be physically expected. But the scatter demonstrates that a broad range of star formation histories can arise, in part due to the realistic variety in possible SFHs that were derived from cosmological simulations (see Section \ref{SEDfittingsect}).}
\label{t10vst90fig}
\end{figure}

It is clear that our two field samples show very similar results at all time intervals, meaning that their results are not sensitive to the details of their definition. We first consider the most recent interval [0,1] Gyr ago (bottom right of Figure \ref{intSFcurvesfig}). Initially we focus on the ancient infallers. Comparing satellite galaxies to the field sample, it is clear that they show reduced star formation. In the ancient infaller panels, there is a clear ordering of the lines from top to bottom (orange, to pink, to red) demonstrating that the reduction in star formation is greater for more massive hosts. Indeed, this pattern is repeated in earlier intervals, and first appears clearly as early as 8-9 Gyr ago, when cluster mass hosts are the first to show a clear deviation from the field samples. By 9-10 Gyr ago, the satellites and field samples are quite similar over most of the stellar mass range.

Now we consider the recent infallers. The amount of suppression the satellites show is always less for the recent infallers compared to the ancient infallers. The mass growth of the recent infallers shows little clear dependence on the host mass. Nevertheless, the presence of some suppression relative to the field sample, albeit smaller, is interesting in itself. Recent infallers have mean infall times of $\sim$1-2.5~Gyr in lookback time. Thus in the [0,1] Gyr interval, this galaxy population is likely to be made of a mixture of objects, some having reached their first pericentre while others may be infalling for the first time, and so it is difficult to determine exactly when the quenching begins. But a significant deviation from the field population appears as early as the [5,6] Gyr interval, which once again suggests quenching begins beyond the virial radius. This is similar to what we saw in the ancient infallers, only now occurring at more recent times, corresponding to the more recent infall of these galaxies.

\subsection{Caveats and tests of systematics}
In Figure \ref{t10vst90fig}, we plot $t_{90}$ (y-axis) versus $t_{10}$ (x-axis) for the relatively isolated galaxies of the field 1 sample, as defined in Section \ref{fieldcompsect}. We colour the data points by the current specific star formation rate. It can be seen that the galaxies tend to separate into two populations, reminiscent of the red sequence and blue cloud, with red points (low sSFR) vertically above blue points (high sSFR) in the figure. Clearly there is some overall trend where the highest values of $t_{90}$ to correspond to the highest values of $t_{10}$. This likely has a very real physical basis, with galaxies that quenched early having accelerated mass growth histories throughout their lives. But the key point is that we see a wide range of possible values of $t_{90}$ at a fixed value of $t_{10}$. This clearly illustrates the wide range of SFHs that the SED fitting tends to recover. As stated in Section \ref{SEDfittingsect}, the fits are performed assuming a large library of diverse SFHs derived from a semi-analytical model. This allows us to span a large parameter space in age and timescale of formation and, as shown in Figure \ref{t10vst90fig}, there is a large variety of $t_{10}$ available in the prior for each $t_{90}$. While an imprint of the priors will always be present in the results, we are positive that the difference between recent and ancient infallers in the old stellar populations is at least in part physical.

It is also worth considering what occurs if a galaxy should suffer strong tidal mass loss of stars. In this case, we are unable to say anything about the stellar population or mass of the stripped stars, as this information is not stored in the SED. Thus the best matching model SFH to the observed SED only describes the mass growth history of the stars that were not tidally stripped, and remain to the current epoch. However, recent results from cosmological simulations have shown that strong stellar mass loss does not arise until more than $\sim$90$\%$ of their dark matter halos have been stripped (\citealp{Smith2016}), and such strong halo mass loss typically requires $>$6~Gyr in their hosts (\citealp{Han2018}). Therefore, in general we do not expect most of the recent infaller population to have suffered significant stellar mass loss. But some of the ancient infallers might suffer some shift horizontally in Figures \ref{diffsfig} and \ref{intSFcurvesfig}, although only in the most recent intervals when time spent in their hosts is sufficiently long. \cite{Han2018} also finds that the more equal the host and satellite are in mass, the more rapid the mass loss of their halos. Thus, at fixed satellite mass, we might expect offsets arising from stellar mass loss to be larger for the lower mass giant hosts than for the more massive cluster hosts. This is the reverse of what is seen in Figure \ref{intSFcurvesfig}, where more massive hosts appear to reduce star formation rates most. Therefore, we conclude that the impact of stellar mass loss on our results is probably not a strong factor governing our results. It is perhaps only slightly cancelling out some of the differences arising from other more significant environmental effects, and this probably occurs only in the ancient infaller panels within the most recent intervals of time.

We also note that the `ancient' and `recent' infallers subsamples actually have a spread in infall times about the mean value. Thus a small fraction of outlier objects in the `ancient infallers' category may not be true ancient infallers, and similarly a small fraction of `recent infallers' may not have recent infall. To consider how this impacts our results we note that, in the extreme case that the two subsamples  were just a random selection of the total sample, then we would not see any differences between the ancient and recent infallers. Thus it is possible that the trends that we see when measuring the differences between ancient and recent infallers (e.g. the `Anc-Rec' column of Figure \ref{tvalfig} or \ref{diffsfig}) would be even stronger if we had a better technique for separating the true ancient and recent infaller. 

Similarly, there are numerous sources of uncertainty present in our data such as on the true velocity dispersion of a host or the true centre of a host. The stellar masses and mass growth history parameters based on SED fitting also have their own intrinsic errors. However, these sources of noise will all tend to make the trends that are visible in the panels of Figures \ref{tvalfig}, \ref{diffsfig} and \ref{intSFcurvesfig} more shallow, which only strengthens the significance of the trends we do see.

\section{Summary  \& Conclusions}
We gather a sample of 27,847 low redshift galaxies ($z$ = 0.018-0.16) that are satellites of central galaxies, based on the \citet{Wang2015} group catalogue. We focus on satellites in the stellar mass range 10$^9$-10$^{11.5}$~M$_\odot$. These inhabit a wide range of host halo masses from 10$^{12}$~M$_\odot$ (massive galaxies) up to 10$^{15}$~M$_\odot$ (massive clusters). By considering both satellite mass and host mass simultaneously, we are better able to unpick the degeneracy between galaxy mass and environment on stellar mass growth history (see for example \citet{Pasquali2010} and \citealp{Peng2010}). 

Next, we plot all of our sample in a single, stacked, projected phase-space diagram. Following the procedures outlined in \citet{Pasquali2019}, we identify which zone in phase-space the galaxies inhabit (see left panel of Figure \ref{compwithsim}). Numerical simulations have shown that with increasing zone number, the mean infall time of the galaxy population in that zone systematically decreases (see right panel of Figure \ref{compwithsim}). Therefore we split our sample into two subsamples; so-called `ancient infallers' (zones $=$ 1 with a mean time since infall of $\sim$5.5~Gyr), and `recent infallers' (zones $\ge$ 5 with a mean time since infall of $\sim$1-2.5~Gyr). By comparing these two subsamples, we can consider the impact of changing the mean infall time on a satellite galaxy population's evolution.

In \citet{Pacifici2016}, near-infrared to ultraviolet SED fitting was conducted to estimate the stellar masses and mass growth histories of a very large sample of galaxies ($\sim$230,000), including all of our sample. We use the parameters provided in this catalogue to constrain the stellar mass growth history of our sample, specifically $t_{10}$, $t_{50}$, and $t_{90}$ (the lookback time when a galaxy forms 10, 50 and 90$\%$ of the total stellar mass formed in its lifetime). We also study and compare the stellar mass growth occurring in one gigayear intervals from the most recent interval [0,1] Gyr ago, back to the most early interval [10,11] Gyr ago.

We are able to simultaneously control for satellite stellar mass and host halo mass, while comparing the differences in stellar mass growth evolution between our `ancient infallers' versus `recent infallers' subsamples. Our main results can be summarised as the following:

\begin{itemize}
\item Galaxies that infall into hosts earlier are quenched earlier. For example, satellites in our ancient infaller sample typically reaches 90$\%$ of their final stellar mass $\sim$2~Gyr earlier than our recent infaller sample. This result is especially clear for our lower mass satellites (10$^{9-10}$~M$_\odot$) in massive hosts ($>$10$^{14}$~M$_\odot$), but the effects are not limited to such objects.
\item By isolating the ancient infaller population using a phase-space analysis, we are more sensitive to the effects of the environment. We can clearly detect the earlier quenching of satellites in hosts with masses of only $\sim$10$^{13}$~M$_\odot$. This mass corresponds roughly to that of low mass groups.
\item The environment of low mass groups is sufficient to alter the stellar mass growth of even quite massive satellites with stellar masses $\sim$10$^{10}$~M$_\odot$).
This highlights how we must better take into account the hierarchical structures in which galaxies tend to be found -- even low mass groups -- to properly understand their evolution.
\item Our ancient infaller sample displays reduced stellar mass growth at epochs as early as $\sim$9~Gyr ago. This is several gigayears before they are expected to enter the virial radius of their eventual host, and at times when their host would be a fraction of its current day mass. Thus it appears the large scale environment surrounding their hosts can already impact their early stellar mass growth.
\end{itemize}

Our results strongly highlight the significance of the substructures in which galaxies are found for their stellar mass growth histories. Hosts with halo masses of less than 10$^{13}$~M$_\odot$, that by some definitions might be considered galaxy hosts rather than groups, can significantly hinder the stellar mass growth of their satellites, even for quite massive satellites. The significance of these low mass groups is also of significance when assembly bias is taken into account. The more dense environment in the outskirts of clusters is more likely to form substructures such as low mass groups, and this may explain why we see reduced stellar mass growth long before the galaxies enter the host in which we see them today.

These results also emphasise the power of our phase-space approach, combined with stellar mass growth histories derived from SED fitting of nearby galaxies. In this way, we attempt to peer back to the early phases of the life of our satellite sample and catch environmentally delayed stellar mass growth, even before they infall into the protocluster version of their host. In the near future we will extend this approach to consider resolved colours, and other galaxy properties. We also see great promise in applying a similar procedure to the satellite population of higher redshift clusters to directly test the predictions derived from the stellar populations of our low redshift sample, and to improve our understanding of the role of environment on galaxy evolution through the early proto-cluster period.

\section*{Acknowledgements}
RS acknowledges support by Sonderforschungsbereich SFB 881 ``The Milky Way System'' (subproject B5) of the German Research Foundation (DFG). P.C-C. was supported by CONICYT (Chile) through Programa Nacional de Becas de Doctorado 2014 folio 21140882.
\noindent

\bibliographystyle{apj}
\bibliography{bibfile}


\section*{Appendix}

\renewcommand\thefigure{A\arabic{figure}}

\setcounter{figure}{0}

\renewcommand\thetable{A\arabic{table}}

\setcounter{table}{0}

\begin{figure*}[h]
\includegraphics[width=180mm]{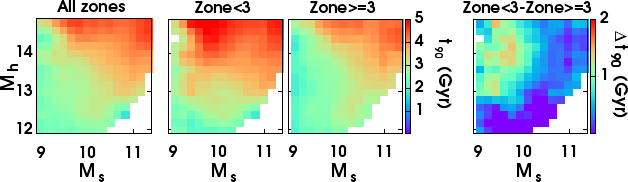}
\caption{These panels are the same as top row of Figure \ref{tvalfig} (looking at values of $t_{90}$ in the M$_{\rm{h}}$-M$_{\rm{s}}$ plane) only this time splitting the sample into a zone $<$ 3 subsample (containing 45$\%$ of the total sample) and a zone $\ge$ 3 subsample (containing the remaining 55$\%$ of the sample). Previously we used a more extreme zone $=$ 1 subsample (referred to as the `Ancient infallers') and a zone $\ge$ 5 subsample (referred to as the `Recent infallers' subsample). The difference between the two subsamples (right panel) is less pronounced than we saw in Figure \ref{tvalfig}, with peak differences being only $\sim$1.2~Gyr (compared to more than 2~Gyr previously). This is as expected, given there is no longer a zone gap between the two subsamples, which in turn means the mean infall times of the two galaxy populations are expected to be more similar (see Figure \ref{compwithsim}). }
\label{tvalfig_diffzones}
\end{figure*}


\begin{figure*}[h]
\includegraphics[width=180mm]{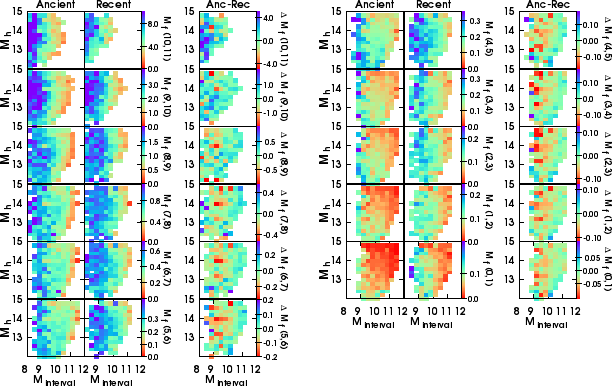}
\caption{Same as caption of Figure \ref{diffsfig} only this time plotting intervals of lookback time measured {\it{from the redshift of the galaxy}} (i.e. t=0 corresponds to the redshift of the galaxy, instead of corresponding to z=0 as it did in Figure \ref{diffsfig}). Our sample of galaxies has a range of redshift from $z$ = 0.018-0.16, therefore the difference in lookback time with this definition compared to previously may be up to 2 Gyr (for the maximum redshift of $z$=0.16). However, there are some advantages to using this definition of lookback time as the mean infall time associated with a zone (see Figure \ref{compwithsim}) is also measured with respect to the galaxy's current redshift. As a result, we can directly see when the changes occur with respect to the mean infall times. These results confirm that the differences between the ancient and recent infallers begin to appear early on, several gigayear prior to the mean infall time of the ancient infaller population, and thus that the stellar mass growth histories are being altered at distances beyond the virial radius of their hosts. However, the broad trends in the panels are fairly similar to Figure \ref{diffsfig}, which is perhaps not surprising given that roughly two-thirds of the sample have $z<$ 0.08 meaning that the interval of the age of the Universe and lookback time differ by less than 1 Gyr.}
\label{diffsfig_lookbackint}
\end{figure*}




\end{document}